\documentclass[11pt]{article}
\usepackage{graphicx} 

\usepackage[utf8]{inputenc}
\usepackage[T1]{fontenc}
\usepackage{lmodern}
\usepackage{geometry}
\usepackage{amssymb}
\usepackage{amsmath}
\usepackage{subcaption}
\usepackage{float}

\usepackage{xspace}

\usepackage{lineno}

\usepackage{titlesec}
\titleformat*{\section}{\bfseries\small}

\usepackage[
  colorlinks=true,        
  linkcolor=blue,          
  urlcolor=blue,           
  citecolor=blue,          
  filecolor=blue,         
  pdfborder={0 0 0} 
]{hyperref}

\usepackage[backend=bibtex, sorting=none, url=false, doi=false]{biblatex}
\AtEveryBibitem{\clearfield{issn}}

\DeclareFieldFormat*{title}{\mkbibquote{#1}}

\renewbibmacro{in:}{}

\DeclareFieldFormat{number}{\textbf{#1}}
\DeclareFieldFormat{pages}{#1}           

\renewbibmacro*{journal+number+pages}{%
  \printfield{journaltitle}%
  \setunit*{\addspace}%
  \printfield{number}%
  \iffieldundef{pages}{}{,\space\printfield{pages}}%
  \setunit*{\addspace}%
  \printtext{(\printfield{year})}%
}

\DeclareBibliographyDriver{article}{%
  \usebibmacro{author/editor+others}%
  \setunit{\labelnamepunct}\newblock
  \usebibmacro{title}%
  \addcomma\space
  \iffieldundef{url}
    {\usebibmacro{journal+number+pages}}
    {\href{\thefield{url}}{\usebibmacro{journal+number+pages}}}%
  \newunit\newblock
  \usebibmacro{finentry}%
}

\DeclareBibliographyDriver{techreport}{%
  \usebibmacro{author/editor+others}%
  \setunit{\labelnamepunct}\newblock
  \usebibmacro{title}%
  \addcomma\space 
  \iffieldundef{number}
    {}
    {%
      \iffieldundef{url}
        {\printfield[number]{number}} 
        {\href{\thefield{url}}{\printfield[number]{number}}}
    }%
  \setunit*{\addspace}%
  \printfield{institution}
  \setunit*{\addspace}%
  \printtext{(\printfield{year})}%
  \newunit\newblock
  \usebibmacro{finentry}%
}

\title{Searches for strong production of supersymmetric particles with the ATLAS detector\\[0.5em]}

\author{Matteo Greco${}^{[1]}$ on behalf of the ATLAS Collaboration\,\footnote{ \textcopyright \ Copyright 2025 CERN for the benefit of the ATLAS Collaboration. CC-BY-4.0 license.}\\
        ${}^{[1]}$\textit{Max-Planck-Institut für Physik, Boltzmannstr. 8, 85748 Garching, Germany}}

\date{Presented at the 32nd International Symposium on Lepton Photon Interactions at High Energies, Madison, Wisconsin, USA, August 25-29, 2025}

\newcommand{\stopone}{$\tilde t_1$}
\newcommand{\cone}{$\tilde c_1$}
\newcommand{\gluino}{$\tilde g$}
\newcommand{\squark}{$\tilde q$}
\newcommand{\stau}{$\tilde \tau$}
\newcommand{\staunu}{$\tilde \nu_\tau$}
\newcommand{\ninone}{$\tilde \chi_1^0$}
\newcommand{\stoponeMass}{$m(\tilde t_1)$}
\newcommand{\coneMass}{$m(\tilde c_1)$}

\newcommand{\ttbar}{$t\bar t$}
\newcommand{\Wt}{$Wt$}
\newcommand{\Wjets}{$W\text{+jets}$}
\newcommand{\Zjets}{$Z\text{+jets}$}

\newcommand*{\pt}{\ensuremath{p_{\text{T}}}\xspace}
\newcommand*{\MET}{\ensuremath{E_{\text{T}}^{\text{miss}}}\xspace}
\newcommand*{\mt}{\ensuremath{m_{\text{T}}}\xspace}

\newcommand{\splitstopone}{$\Delta m(\tilde t_1, \tilde \chi_1^0)$}
\newcommand{\MassPlaneStopone}{$m(\tilde t_1)-m(\tilde \chi_1^0)$}

\begin{document}

\maketitle
\vspace{-0.8cm}
\begin{abstract}
    Supersymmetry (SUSY) provides elegant solutions to several open questions in the Standard Model, and searches for SUSY particles are an important component of the LHC physics program. Naturalness arguments favour supersymmetric partners of the gluons and third-generation quarks with masses light enough to be produced at the LHC. With increasing mass bounds on more classical Minimal Supersymmetric Standard Model (MSSM) scenarios other variations of supersymmetry, including non-minimal particle content, become increasingly interesting. This proceeding will present the latest results of searches conducted by the ATLAS experiment at LHC at center of mass energies of $\sqrt{s}=13$ and 13.6~TeV which target gluino and squark production, including stop, in a variety of decay modes.
\end{abstract}

\section{Introduction}

The Standard Model of Particle Physics (SM), despite its detailed and precise formulation and the large number of experimental confirmations, still has several unresolved open questions, such as the hierarchy problem, the lack of unification of the fundamental interactions, and the absence of an explanation for dark matter. Among the several beyond Standard Model (BSM) theories, one of the most promising to solve these problems is \textit{Supersymmetry} (SUSY). This theory foresees for each SM particle the existence of an associated \textit{supersymmetric partner} (or \textit{superpartner}) with the same quantum numbers but with spin (S) differing by one-half: for each quark and lepton the associated scalar superpartners are called \textit{squarks} ($\tilde q$) and \textit{sleptons} ($\tilde \ell$) respectively; for each gauge boson the associated fermionic superpartners are \textit{gauginos} (\textit{zino}, $\tilde Z$, \textit{wino}, $\tilde W$, \textit{photino}, $\tilde \gamma$, and \textit{gluino}, $\tilde g$); and for the Higgs sector (two Higgs doublets are foreseen) the associated fermionic superpartners are called \textit{higgsinos} ($\tilde H$). The simplest formulation of SUSY is called \textit{Minimal Supersymmetric Standard Model} (MSSM), where baryonic (B) and leptonic (L) numbers are not conserved separately, and the \textit{R-parity} quantum number, defined as R-parity~=~$(-1)^{\text{3(B}-\text{L)+2S}}$, and its conservation are introduced. In MSSM, gauginos and higgsinos mix together to form four charged particles called \textit{charginos} ($\tilde\chi_i^\pm$, with $i$=1,2) and four neutral particles called \textit{neutralinos} ($\tilde\chi_j^0$, with $j$=1,2,3,4) and if R-parity is conserved the \ninone~ represents the \textit{Lightest Supersymmetric Particle} (LSP), which is stable and a good dark matter candidate. Supersymmetric particles produced in strong interactions (squarks and gluinos) have significantly larger production-cross section than non-colored sparticles of equal masses, and so far searches performed by the ATLAS collaboration \cite{ATLAS-col} have set very stringent limits on the top-squark (\stopone, also called \textit{stop}) mass and production cross-section.\\
Some of the latest results published by the ATLAS Collaboration are presented here, involving three searches~\cite{ttMET-paper, tcMET-paper, ccMET-paper} targeting top-squark production, based on proton–proton collision data at a center of mass energy $\sqrt{s}=13$~TeV collected during Run~2 of the LHC, covering the data-taking period from 2015 to 2018 (corresponding to an integrated luminosity of 139-$140$~fb$^{-1}$). In addition, one search~\cite{SUSYtaus-paper} for gluino and squark production is discussed, which analyzes both Run~2 data and a portion of the Run~3 dataset, collected between 2022 and 2023, the latter corresponding to $\sqrt{s}=13.6$~TeV and to an integrated luminosity of $51.8$~fb$^{-1}$.

\section{Search for \stopone\stopone~pairs in $tt$+\MET~final states}

\begin{figure}[!b]
\begin{minipage}{0.5\textwidth}
    \includegraphics[width=0.95\linewidth]{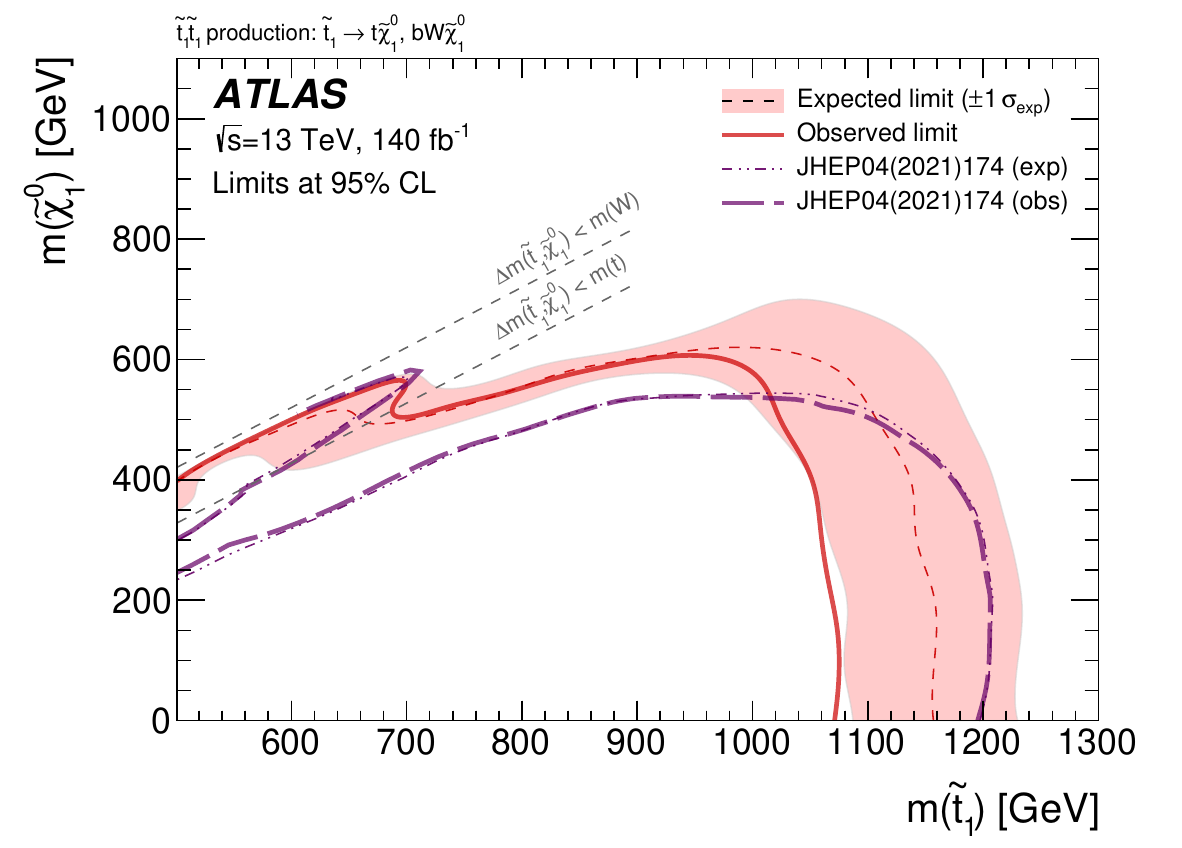}
    \subcaption{}
\end{minipage}
\begin{minipage}{0.5\textwidth}
    \includegraphics[width=0.95\linewidth]{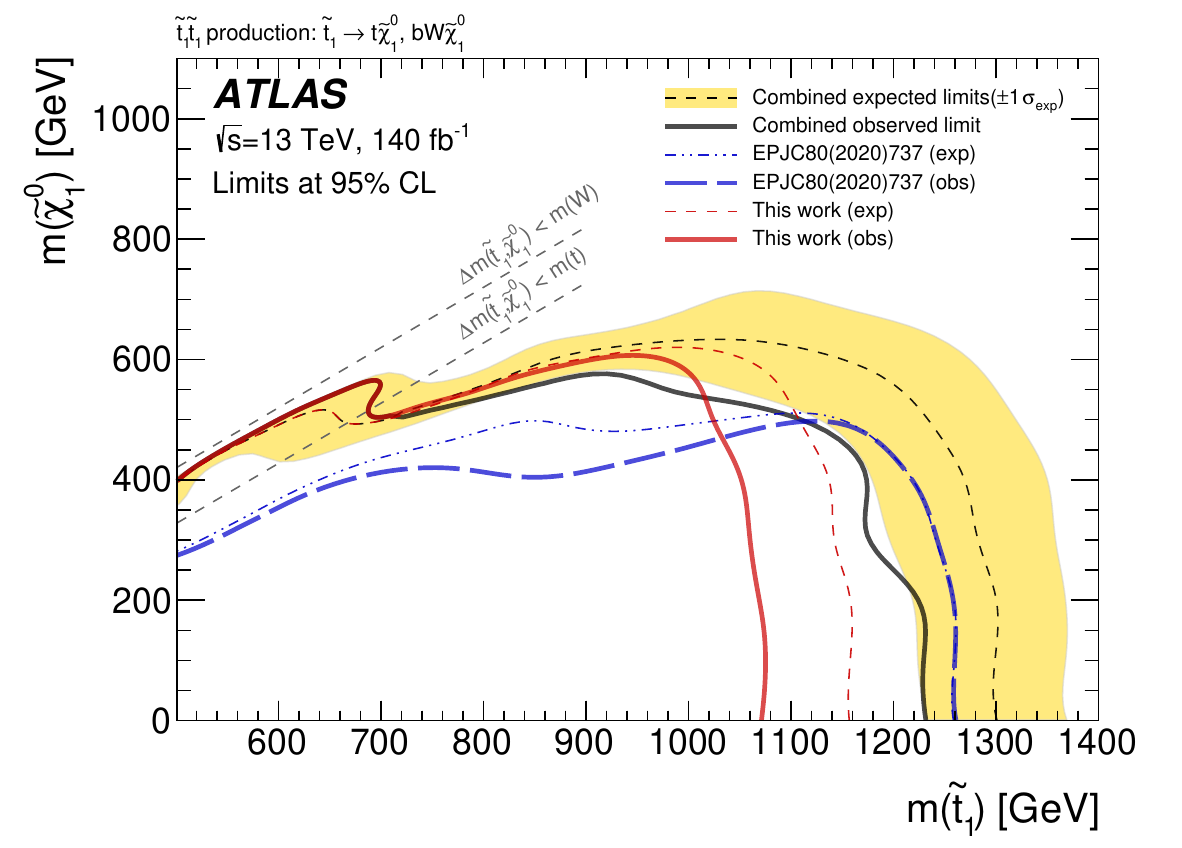}
    \subcaption{}
\end{minipage}
\caption{Expected and observed exclusion limits at 95\% CL in the \MassPlaneStopone~plane: (a) shows a comparison between the results of the described analysis and those of a single-lepton search~\cite{ATLAS2021-1Lanalysis}, while (b) shows the results of this analysis, those of a zero-lepton analysis~\cite{indep-0L} and their combination~\cite{ttMET-paper}.}
\label{fig:ttMET}
\end{figure}

The first presented analysis~\cite{ttMET-paper} consists of a search for top-squark pair (\stopone\stopone) production decaying into a top-quark pair (\ttbar) and two LSPs (\ninone), where one top-quark decays leptonically ($t\rightarrow b\ell\nu$) and the other hadronically ($t\rightarrow bq\bar q$). Only two top-squark decay models are considered according to the values of the mass difference (\textit{mass-splitting}) between the \stopone~and \ninone~masses, \splitstopone: the \textit{two-body} decay mode, if the mass-splitting is larger than the top-quark mass, and the \textit{three-body} decay mode, if the value is between the mass of the $W$-boson and the mass of the top-quark. The final state of interest is composed of one isolated charged lepton, jets and missing transverse energy (\MET). \\
The analysis strategy focuses on two mutually exclusive categories: a \textit{resolved high-\MET} category, targeting regions with resolved top-quark decays, meaning low-\pt~top-quarks (i.e. \pt~$\in[200, 600]$~GeV), an absence of large-R jets and large values of the missing transverse energy magnitude (\MET~$>230$~GeV); and a \textit{boosted} category, looking instead at regions with boosted top-quark decays, meaning high-\pt~top-quarks (i.e. \pt~$>600$~GeV) and at least one large-R jet. The signal sensitivity is enhanced in properly defined signal regions (SRs). For the high-\MET~category, SRs are defined based on the number of $b$-tagged jets, namely jets originating from $b$-quarks, also referred to as $b$-jets, in the events. For the boosted category, signal events are required to have at least one large-R jet, and are first classified according to whether the leading-\pt large-R jet is top-tagged or not, and afterwards based on the number of $b$-jets inside and outside the large-R jet. A neural-network (NN) is exploited to further improve the signal/background discrimination.\\
The main backgrounds affecting the analysis are \ttbar~production (either semileptonic (1L), with at least one prompt lepton, or fully leptonic (2L), with at least two prompt leptons), single-top (including \Wt~production) and \Wjets. Their estimation is obtained from specific control regions (CRs). In order to discriminate \ttbar~and \Wjets~processes, the product of the transverse mass \mt\footnote{The transverse mass \mt is the transverse mass of the lepton \pt and the missing transverse momentum.} and the charge of the lepton $q(\ell)$ is used: \mt has a kinematic endpoint at the mass of the $W$-boson for \ttbar-1L and \Wjets, while the endpoint is much higher for \ttbar+2L, whereas $q(\ell)$ helps in discriminating \ttbar-1L and \Wjets~because of the charge asymmetry of the latter.\\ 
The background estimations are extrapolated to the SRs, where no significant excess of events is observed. Therefore expected and observed exclusion limits at 95\% of confidence level (CL) are calculated. As shown in \figurename~\ref{fig:ttMET} (a) the results significantly improve the sensitivity in the region with small mass-splitting values, with respect to the previous limits imposed by a single-lepton analysis~\cite{ATLAS2021-1Lanalysis}. The results are also combined with those obtained by an independent analysis looking at a zero-lepton channel~\cite{indep-0L}, as shown in \figurename~\ref{fig:ttMET} (b), excluding top-squark masses up to 1230~GeV (for massless \ninone) and neutralino masses up to 570~GeV.

\section{Search for \stopone\stopone~pairs in $tc$+\MET~final states}
\label{sec:tcMET}

\begin{figure}[!h]
\begin{minipage}{0.5\textwidth}
    \includegraphics[width=0.95\linewidth]{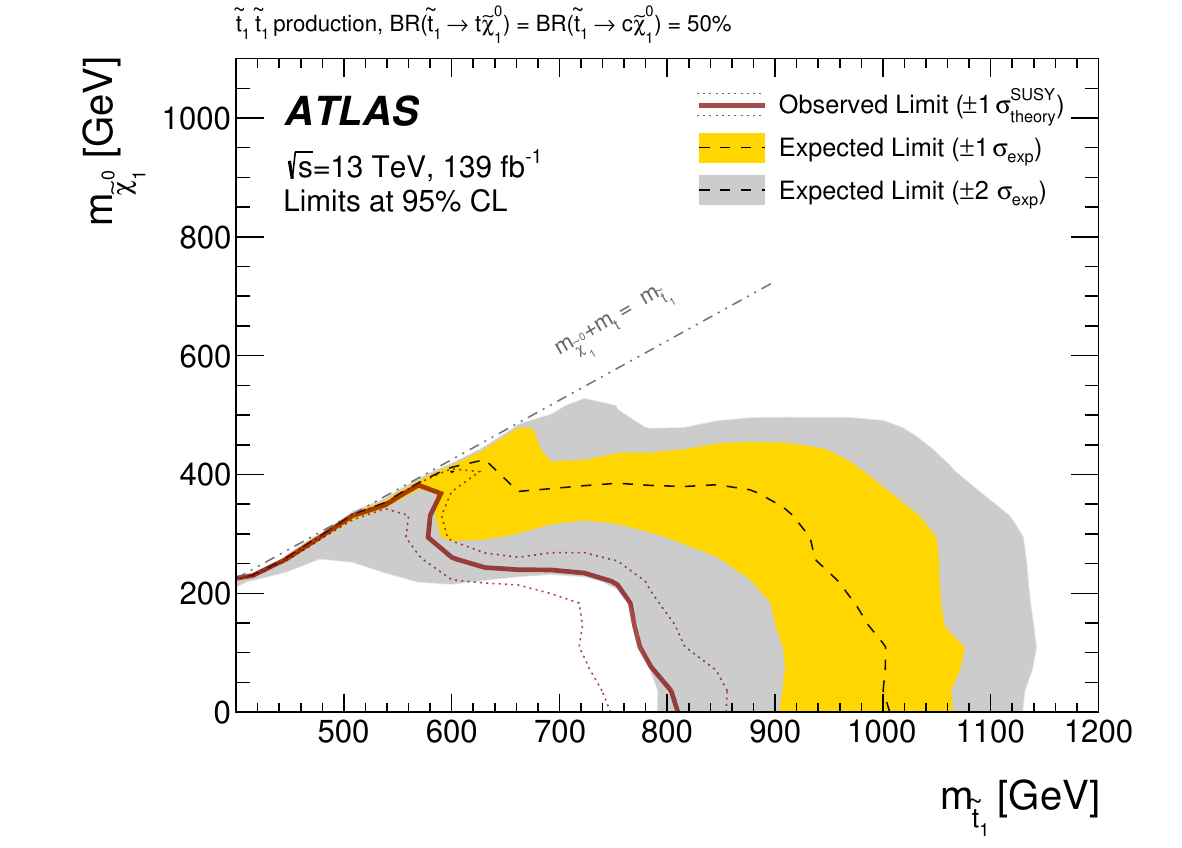}
    \subcaption{}
\end{minipage}
\begin{minipage}{0.5\textwidth}
    \includegraphics[width=0.95\linewidth]{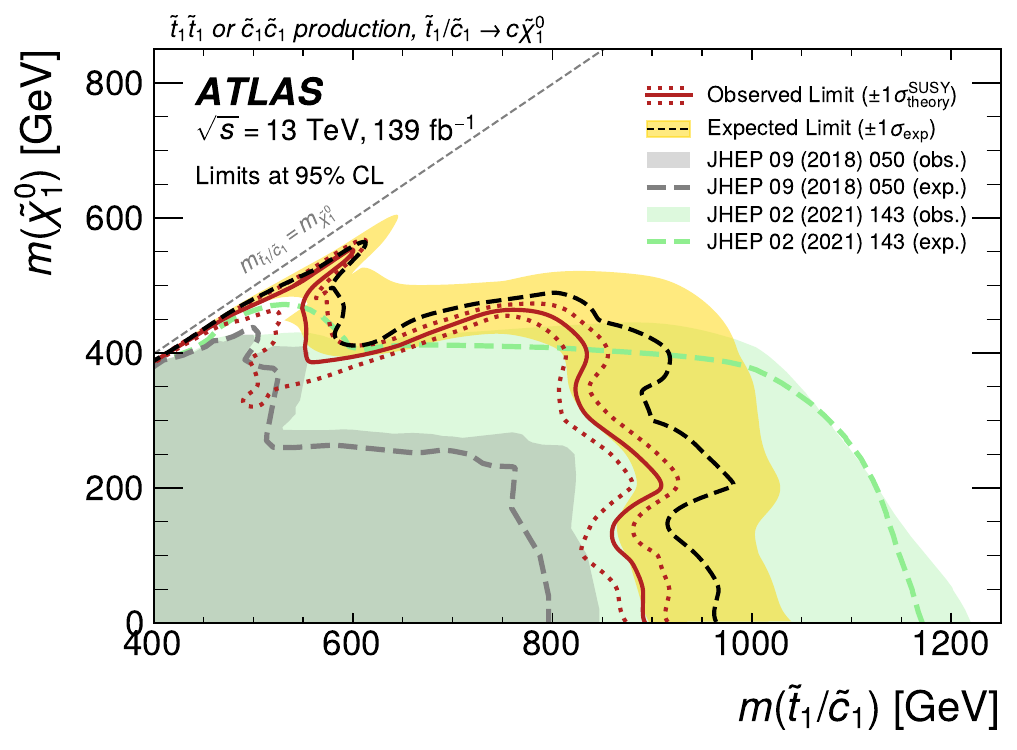}
    \subcaption{}
\end{minipage}
\caption{Expected and observed exclusion limits at 95\% CL in (a) the \MassPlaneStopone~plane for the search of top-squark pairs decaying in $tc$+\MET~final states \cite{tcMET-paper} and in (b) the $m($\stopone/\cone$)-m($\ninone$)$ plane for the search of top or charm-squark pairs decaying in $cc$+\MET~final states \cite{ccMET-paper}.}\label{tcMETandccMET}
\end{figure}

The second analysis \cite{tcMET-paper} is dedicated to the search for top-squark pair production with equally probable decays \stopone~$\rightarrow t$\ninone~ or \stopone~$\rightarrow c$\ninone~ in hadronic final states. This particular channel slightly relaxes the constraints on the MSSM allowing for an extension of it featuring a \textit{non-minimal flavour violation} (MFV) in the II and III generations of the squark sector. The final state is composed of one $b$-tagged jet, one $c$-tagged jet, which is a jet originating from $c$-quark, also referred to as $c$-jet, an absence of leptons and large \MET (\MET>~230~GeV). Only the phase-space regions with large mass splittings (\splitstopone$\gtrsim175$~GeV) are investigated. The advancement in the $c$-tagging jet techniques in ATLAS has motivated the exploration of final states with $c$-quarks: a $c$-tagging jet algorithm based on the $b$-tagging jet DL1r algorithm~\cite{DL1r-ref} but with tuned parameters optimized for identifying jets that contain $c$-jets is developed.\\
The analysis strategy consists of dividing the parameter space into three different regions according to the mass-splitting values: \textit{bulk}, for values much greater than the top-quark mass (\splitstopone~$\gg m(t)$), \textit{intermediate} with intermediary values (\splitstopone~$> m(t)$) and further divided into `boosted' and `resolved' top-quark cases, and \textit{compressed} with values of the order of the top-quark mass (\splitstopone~$\sim m(t)$). Signal regions are defined for each of these categories exploiting properties of different kinematic variables. In particular, in the compressed scenario, the signal sensitivity is enhanced by requiring the presence of an \textit{initial state radiation} (ISR) jet (neither $b$ nor $c$-tagged) and training a NN. The relevant backgrounds come from \Zjets~processes (estimated in CRs using events with two leptons) and single-top, \ttbar~and \Wjets~processes (estimated in CRs containing events with one lepton).\\
No significant excess of events is observed in any of the defined SRs and the 95\% CL exclusion limits (see \figurename~\ref{tcMETandccMET} (a)) show that top-squarks can be excluded for masses up to 800~GeV (for massless \ninone) and 600~GeV in the compressed scenario. This analysis is also reinterpreted into a search~\cite{ATL-PHYS-PUB-2025-010} for \textit{flavour-violating dark matter models} excluding the existence of new mediators coupling with dark matter particles and SM quarks for masses up to 1.2~TeV at 95\% CL.

\section{Search for \stopone\stopone~and \cone\cone~pairs in $cc$+\MET~final states}

The third search presented~\cite{ccMET-paper} targets the \stopone\stopone~and \cone\cone~pair production, both via the decay of \stopone/\cone~$\rightarrow c$\ninone. The analysis features SUSY models with minimal flavour violating effect, for which charm-squarks could be considered lighter than other squarks. The final state of interest is composed of two $c$-tagged jets, no leptons and missing transverse energy.
Two different kinematic regions are investigated: \textit{high-mass}, sensitive to SUSY models with \stoponeMass, \coneMass~$\gtrsim 600$~GeV and $\Delta m($\stopone/\cone,\ninone$)\gtrsim 200$~GeV, and \textit{compressed}, targeting smaller squark masses and $\Delta m($\stopone/\cone,\ninone$)< 175$~GeV. For the high-mass region the leading jet is required to be $c$-tagged and \MET-based variables are used to improve the signal/background discrimination. In the compressed region, instead, the sensitivity is mainly enhanced by requiring high momentum jets from ISR and using \textit{recursive jigsaw reconstruction} (RJR)~\cite{PhysRevD.96.112007}, while requiring the leading jet to not be $c$-tagged. Like the analysis discussed in Section~\ref{sec:tcMET}, the dominant backgrounds arising from \Zjets~and \Wjets~processes are estimated using similar methods.\\
No significant excess of data over the expected SM background is observed. The obtained expected and observed exclusion limits are reported in \figurename~\ref{tcMETandccMET} (b) showing that top/charm-squark masses can be excluded up to about 900~GeV, in the hypothesis of massless neutralinos at 95\% CL. Furthermore, this result improves  the previous limits~\cite{Aaboud2018} imposed by ATLAS by about 100~GeV, while being complementary to a single-squark search not using $c$-tagging \cite{Aad2021}.

\section{Gluinos and squarks pair production}

\begin{figure}[!h]
\begin{minipage}{0.5\textwidth}
    \includegraphics[width=0.95\linewidth]{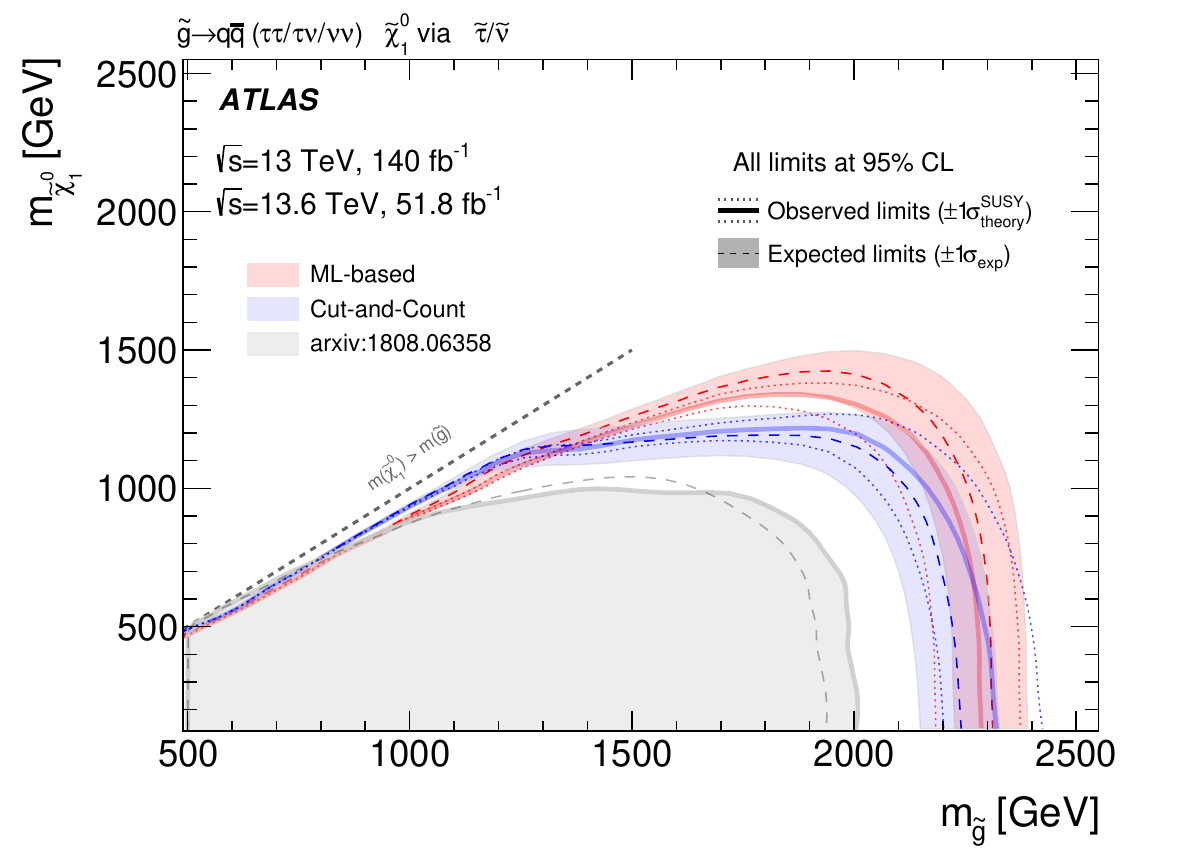}
    \subcaption{}
\end{minipage}
\begin{minipage}{0.5\textwidth}
    \includegraphics[width=0.95\linewidth]{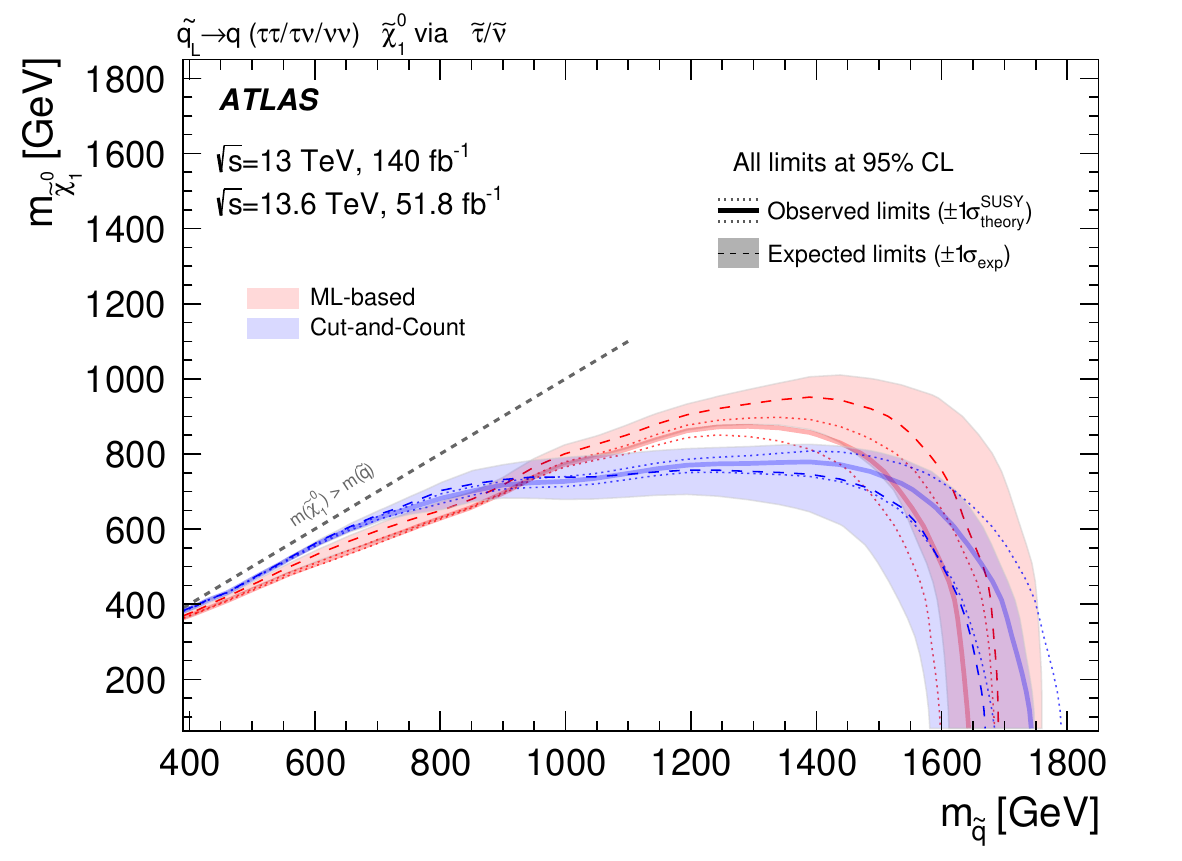}
    \subcaption{}
\end{minipage}
\caption{Expected and observed exclusion limits at 95\% CL in (a) $m(\tilde g)-m($\ninone) plane and (b) $m(\tilde q)-m($\ninone) plane for the search of \gluino\gluino~and \squark\squark~pair productions respectively~\cite{SUSYtaus-paper}.}\label{fig:gluino}
\end{figure}

The last analysis presented here~\cite{SUSYtaus-paper} is a search for the pair production of gluinos (\gluino\gluino) and either I or II generation left-handed squarks (\squark\squark) with $\tau$-sleptons (\stau) and $\tau$-sneutrinos (\staunu), using Run~2 and partial Run~3 data. The final state of interest consists of $\tau$-leptons, jets and missing transverse energy. In particular three different channels are considered according to the number of hadronically-decaying $\tau$-leptons and leptons present in the event: one hadronically-decaying $\tau$ and no leptons (\textit{1tau0lep}), one hadronically-decaying $\tau$ and one electron/muon (\textit{1tau1lep}) and two hadronically-decaying $\tau$'s (\textit{2tau}). For each of the considered channels two different approaches are adopted: a traditional \textit{cut-and-count}, improving the signal sensitivity defining selections using \MET-related variables, and a \textit{machine-learning} approach, performing a multiclass classification with boosted decision trees (BDTs), which classify events into signal and various background categories, including \Wjets, \Zjets, diboson, top and fake-$\tau$ processes. For each of the three channels the BDTs are trained using a five-fold cross-validation and used for defining control, validation and signal regions in the analysis.\\
For both the adopted approaches no significant deviations from the SM expectations are observed in data. Exclusion limits for gluino and squark models are reported in \figurename s~\ref{fig:gluino} (a) and (b) respectively for both the cut-and-count and machine-learning methods. Gluino masses can be excluded up to 2.25~TeV and neutralino up to 1.35 TeV (for gluino masses of $\sim2$~TeV) at 95\% CL. Squarks masses are excluded up to 1.7~TeV and neutralino masses up to 0.85~TeV at 95\% CL. The cut-and-count strategy performs better with respect to the machine-learning one in the compressed phase-space region because of a larger number of events and dedicated selections optimized for improving sensitivity in that region, while for higher gluino/squark masses the machine-learning approach is more sensitive than the cut-and-count and even more for higher neutralino masses.

\section*{Conclusions}
Some of the latest results conducted by the ATLAS collaboration at LHC searching for supersymmetric particles produced in strong interactions, by using proton-proton collision data of LHC collected during Run~2 and partially during Run~3, are presented here, mainly looking at top-squark, charm-squark and gluino pair productions. In none of the analyses significant excesses of events are observed, thus exclusion limits are placed in terms of SUSY particle masses.

\printbibliography

\end{document}